\documentclass[journal]{IEEEtran}

\usepackage{cite}
\usepackage{amsmath,amssymb,amsfonts}
\usepackage{algorithmic}
\usepackage{graphicx}
\usepackage{setspace}
\usepackage{textcomp}
\usepackage{bm}
\usepackage{array}
\usepackage{url}
\usepackage{xcolor}
\usepackage{comment}
\usepackage{fancyhdr}

\makeatletter
\let\MYcaption\@makecaption
\makeatother

\usepackage{subcaption}
\makeatletter
\let\@makecaption\MYcaption
\makeatother

\newcommand{\1}{\mbox{1}\hspace{-0.25em}\mbox{l}}
\ifCLASSINFOpdf
\else
\fi
\newcommand{\MYheader}{\smash{\scriptsize
\hfil\parbox[t][\height][t]{\textwidth}{\centering
This paper has been published in IEEE Journal of Biomedical and Health Informatics (J-BHI).
Digital Object Identifier (DOI) is \protect\url{https://doi.org/10.1109/JBHI.2023.3345486}}\hfil\hbox{}}}

\newcommand{\MYfooter}{\smash{\scriptsize
\hfil\parbox[t][\height][t]{\textwidth}{\centering
© 2024 IEEE.  Personal use of this material is permitted.  Permission from IEEE must be obtained for all other uses, in any current or future media, including reprinting/republishing this material for advertising or promotional purposes, creating new collective works, for resale or redistribution to servers or lists, or reuse of any copyrighted component of this work in other works.}\hfil\hbox{}}}

\makeatletter

\def\ps@headings{%
\def\@oddhead{\mbox{}\scriptsize\MYheader \hfil \thepage}
\def\@evenhead{\scriptsize\thepage \hfil \MYheader\mbox{}}
\def\@oddfoot{\MYfooter}%
\def\@evenfoot{\MYfooter}}

\def\ps@IEEEtitlepagestyle{%
\def\@oddhead{\mbox{}\scriptsize\MYheader \hfil \thepage}%
\def\@evenhead{\scriptsize\thepage \hfil \MYheader\mbox{}}%
\def\@oddfoot{\MYfooter}%
\def\@evenfoot{\MYfooter}}

\makeatother

\begin{document}
%
\title{CalibrationPhys: Self-supervised Video-based \\Heart and Respiratory Rate Measurements by \\Calibrating Between Multiple Cameras}
%
%
%

\author{Yusuke Akamatsu, Terumi Umematsu, and Hitoshi Imaoka
\thanks{Yusuke Akamatsu, Terumi Umematsu, and Hitoshi Imaoka are with Biometrics Research Laboratories, NEC Corporation, Kawasaki, Japan (e-mail: yusuke-akamatsu@nec.com; terumi@nec.com; h-imaoka\_cb@nec.com). }
}

%
%

\markboth{Journal of \LaTeX\ Class Files,~Vol.~XX, No.~XX, July~2023}%
{Shell \MakeLowercase{\textit{et al.}}: Bare Demo of IEEEtran.cls for IEEE Journals}
%



\maketitle


\begin{abstract}
Video-based heart and respiratory rate measurements using facial videos are more useful and user-friendly than traditional contact-based sensors.
However, most of the current deep learning approaches require ground-truth pulse and respiratory waves for model training, which are expensive to collect.
In this paper, we propose CalibrationPhys, a self-supervised video-based heart and respiratory rate measurement method that calibrates between multiple cameras.
CalibrationPhys trains deep learning models without supervised labels by using facial videos captured simultaneously by multiple cameras.
Contrastive learning is performed so that the pulse and respiratory waves predicted from the synchronized videos using multiple cameras are positive and those from different videos are negative.
CalibrationPhys also improves the robustness of the models by means of a data augmentation technique and successfully leverages a pre-trained model for a particular camera.
Experimental results utilizing two datasets demonstrate that CalibrationPhys outperforms state-of-the-art heart and respiratory rate measurement methods.
Since we optimize camera-specific models using only videos from multiple cameras, our approach makes it easy to use arbitrary cameras for heart and respiratory rate measurements.
\end{abstract}


\section{Introduction}
\label{sec:introduction}
\IEEEPARstart{V}{ital} sign monitoring is essential in various fields spanning the medical, healthcare, sports, emotion recognition, and driver monitoring areas.
Heart rate (HR) and respiratory rate (RR) are traditionally measured by skin-contact sensors such as photoplethysmography (PPG) and respiratory belts, respectively.
However, contact sensors may cause discomfort when worn and increase the risk of infection because the sensors are shared with others.
Measurement devices such as pulse oximeters and wearable sensors are also required.
Recently, the utilization of computers and smartphones to measure HR and RR from facial videos has been attracting interest.
This video-based approach enables the non-contact measurement of vital signs and remote vital monitoring for applications such as online meetings and telemedicine.

For HR measurement, pulse waves are extracted by capturing subtle changes in reflected light on the skin, which is caused by changes in the volume of blood vessels with the cardiac cycle~\cite{poh2010non}.
HR estimation methods first acquire the red, green, and blue channel (RGB) signals from a facial region of interest (ROI) by capturing the face with a camera.
Then, to separate the pulse waves from the noise, signal processing approaches such as blind source separation~\cite{poh2010non,poh2010advancements} or skin reflection models~\cite{de2013robust,wang2016algorithmic} are adopted.
For RR measurement, the chest and head movements associated with respiratory motion, $i.e.,$ respiratory waves, are extracted.
RR estimation methods obtain pixel movement based on optical flow~\cite{li2014non,wiede2017remote,gwak2022motion} or pixel intensity variation~\cite{bartula2013camera,tarassenko2014non,massaroni2019non} from chest or facial ROIs.
Zhan {\it et al.}~\cite{zhan2020revisiting} showed that pixel movement-based methods perform better RR estimation than pixel intensity variation-based methods.
These signal processing approaches extract handcrafted features based on certain assumptions about HR and RR.
However, these assumptions may not hold in unconstrained scenarios ($e.g.$, severe body motion and illumination changes), resulting in poor estimation performance.

\begin{figure*}[t]
\begin{center}
\includegraphics[scale=0.4]{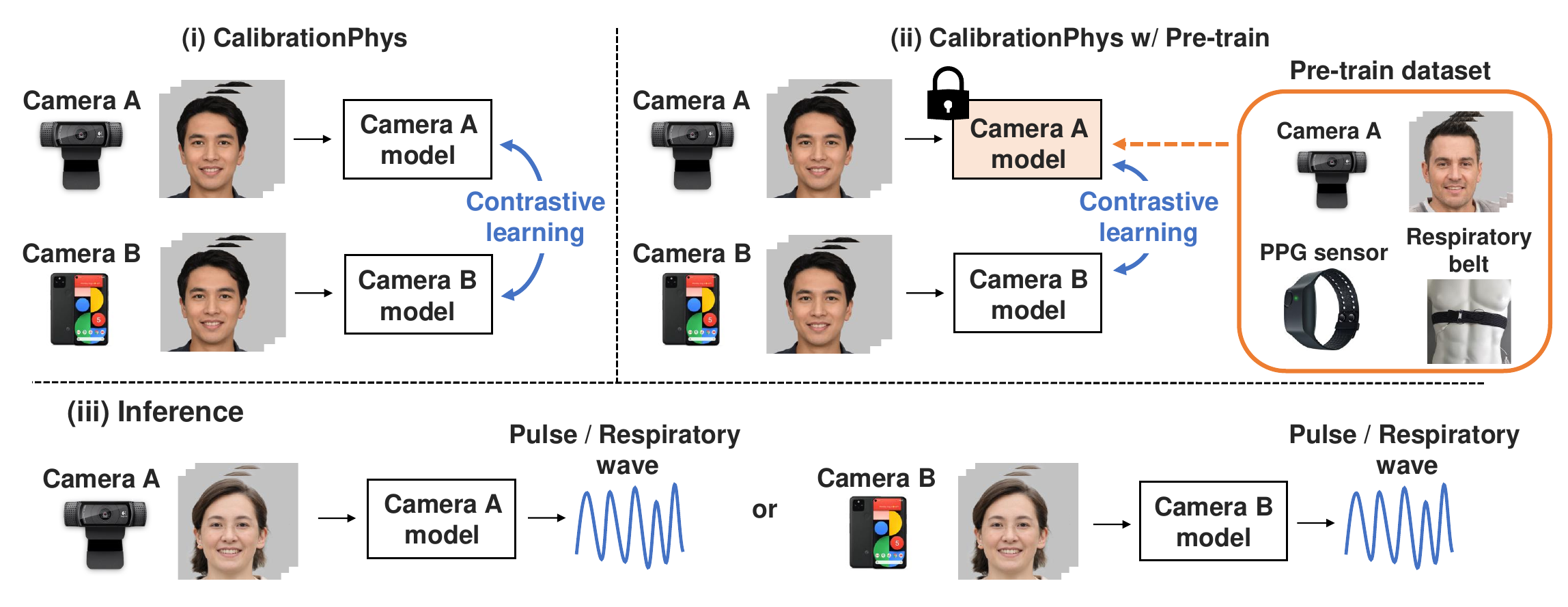}
\end{center}
\caption{(i) CalibrationPhys performs contrastive learning using facial videos captured simultaneously by two cameras A and B (a webcam and a smartphone in this case) to train HR and RR estimation models for each camera. (ii) CalibrationPhys w/ Pre-train uses the pre-trained HR or RR estimation model for a particular camera (the webcam in this case). In contrastive learning, we fix the pre-trained model and train only the model for the newly applied camera (the smartphone in this case). (iii) During inference, we predict pulse and respiratory waves from the facial videos by using the trained model for each camera. In our experiments, we used a Logitech C920n HD PRO webcam \protect \footnotemark[1] (Camera A), a Google Pixel~5 smartphone \protect \footnotemark[2] (Camera B), the wearable sensor E4 wristband~\protect \footnotemark[3] (PPG sensor), and the RIP respiratory sensor \protect \footnotemark[4] (Respiratory belt). Photos by Generated.Photos. \protect \footnotemark[5]}
\label{fig:intro}
\end{figure*}

To achieve robust estimation in various scenarios, deep learning (DL)-based HR~\cite{chen2018deepphys,niu2018synrhythm,niu2019rhythmnet,yu2019remote,niu2019robust,liu2020multi,yu2022physformer,liu2023efficientphys} and RR~\cite{chen2018deepphys,liu2020multi,zhan2020revisiting} estimation methods have been proposed.
These methods train 2D convolutional neural networks (2DCNN)~\cite{chen2018deepphys,liu2020multi,liu2023efficientphys}, 3DCNN~\cite{yu2019remote}, and transformer~\cite{yu2022physformer,liu2023efficientphys} so that the pulse or respiratory waves can be predicted from the facial video.
However, these methods require a large-scale dataset that includes facial videos and ground-truth pulse and respiratory waves, which are used as supervised labels.
Although facial videos can be collected relatively easily, it is expensive to collect pulse and respiratory waves using pulse sensors and respiratory belts.
Data collection wearing these devices is not only a burden on the subject but also on the data collector, who must strictly synchronize the videos with the devices.
Furthermore, the need for supervised labels causes considerable problems when cameras are changed.
Specifically, the estimation performance of HR and RR depends on the camera, since each camera has different colors and internal image processing.
Therefore, when a camera is changed, facial videos and supervised labels should ideally be re-collected in order to train a specific DL model for that camera.
However, collecting facial videos and supervised labels for each camera requires much time and effort, so the use of arbitrary cameras is still difficult.

In this paper, we propose a self-supervised video-based HR and RR measurement method called {\it CalibrationPhys} that calibrates between multiple cameras.
The proposed method trains HR and RR estimation models without supervised labels by using videos captured simultaneously with multiple cameras ($e.g.,$ webcam and smartphone).
Since training is performed using only the videos from multiple cameras, it is easy to build a DL model when the camera is changed.
Figure~\ref{fig:intro} shows an overview of the proposed method.
CalibrationPhys optimizes the HR and RR estimation models for each camera by means of contrastive learning utilizing facial videos captured by the two cameras (see Fig.~\ref{fig:intro}~(i)).
In contrastive learning, we assume that the HR and RR estimated from the facial videos captured simultaneously by the two cameras are similar, and that their values estimated from the different facial videos are not similar.
We further introduce a data augmentation method for HR and RR to make contrastive learning more effective and improve the robustness of the estimation models.

We can also consider another case here: the leverage of a pre-trained model.
Specifically, if we have already trained HR or RR estimation models for a particular camera, we can reuse them as pre-trained models, as there is a possibility that the facial videos and supervised labels for a particular camera have already been collected.
In this case, we fix the pre-trained model and train only the model for the new camera.
We call this CalibrationPhys w/ Pre-train, as shown in Fig.~\ref{fig:intro}~(ii).
CalibrationPhys w/ Pre-train can be viewed as a kind of domain adaptation utilizing one pre-trained camera ($i.e.,$ source camera) and one newly applied camera ($i.e.,$ target camera).
During inference, the optimized models predict the pulse and respiratory waves from the facial video (see Fig.~\ref{fig:intro}~(iii)), and then we calculate HR and RR from the waves.
Note that only a single camera is needed for inference.
Experiments on two datasets demonstrate that the proposed method outperforms state-of-the-art HR and RR estimation methods.
Our main contributions are as follows.
\begin{itemize}
\item We present the first self-supervised DL-based method trained using facial videos captured by two RGB cameras. This approach does not require supervised labels and makes it easy to use arbitrary cameras. We can apply this framework to both HR and RR measurements.
\item We introduce a data augmentation technique called {\it temporal augmentation}. This technique expands the original limited training dataset and improves the robustness of the HR and RR measurements.
\item Few studies have been conducted on the differences in HR and RR estimation performance between different cameras.
We investigate the performance differences between a webcam and a smartphone.
\item Through experiments on a private multi-camera dataset and a public UBFC-rPPG~\cite{bobbia2019unsupervised}, we confirmed that the proposed method is superior to state-of-the-art methods in both estimation performance and computational cost.
\end{itemize}

\footnotetext[1]{\url{https://www.logicool.co.jp/ja-jp/products/webcams/hd-pro-webcam-c920n.960-001261.html}}
\footnotetext[2]{\url{https://store.google.com/gb/category/phones?hl=en-GB}}
\footnotetext[3]{\url{https://www.empatica.com/research/e4/}}
\footnotetext[4]{\url{https://www.creact.co.jp/measure/bio/biosignalsplux/file/RIP-Datasheet.pdf}}
\footnotetext[5]{\url{https://generated.photos/}}

\section{Related Work}
\label{sec:related}
This section reviews recent work in two key areas related
to this paper: supervised DL-based HR and RR measurement (\ref{subsec:supervised}) and self-supervised DL-based HR measurement (\ref{subsec:self-supervised}).

\subsection{Supervised DL-based HR and RR Measurement}
\label{subsec:supervised}
Several DL-based approaches have been proposed for video-based HR~\cite{chen2018deepphys,niu2018synrhythm,niu2019rhythmnet,yu2019remote,niu2019robust,liu2020multi,yu2022physformer,liu2023efficientphys} and RR~\cite{chen2018deepphys,liu2020multi,zhan2020revisiting}  measurements.
DeepPhys~\cite{chen2018deepphys} is a VGG-style 2DCNN that predicts pulse or respiratory waves from facial videos.
DeepPhys is composed of two models: a motion model, where the normalized video frame differences are input, and an appearance model, where the motion representation is captured by an attention mechanism.
TS-CAN~\cite{liu2020multi} incorporates a temporal shift module (TSM)~\cite{lin2019tsm} into DeepPhys to capture temporal information and thereby improve HR and RR estimation performance.
The authors also proposed MTTS-CAN, which predicts both pulse and respiratory waves in a multi-task manner, but its performance is lower than that of TS-CAN.
EfficientPhys~\cite{liu2023efficientphys} is a preprocessing-free version of TS-CAN that efficiently predicts pulse waves from facial videos.
The authors proposed a convolution-based and a transformer-based EfficientPhys and showed that the convolution-based model performs better HR estimation.
While DeepPhys, TS-CAN, and EfficientPhys take two consecutive video frames as input, another type of DL-based method~\cite{niu2018synrhythm,niu2019rhythmnet,yu2019remote,niu2019robust,yu2022physformer} inputs a spatio-temporal representation to its network.
PhysNet~\cite{yu2019remote} is a 3DCNN-based spatio-temporal network that predicts pulse waves from facial videos.
It can provide a more comprehensive representation of the spatial and temporal information of the physiological signals in the video stream~\cite{cheng2021deep}.
However, these supervised HR and RR measurement methods require supervised labels such as pulse and respiratory waves.
Collecting these supervised labels is not easy, so the training cost of the estimation model is still expensive.

\subsection{Self-supervised DL-based HR Measurement}
\label{subsec:self-supervised}
To overcome the problem of supervised methods, recent HR estimation methods focus on self-supervised (or unsupervised) learning approaches~\cite{gideon2021way,yue2022video,wang2022self,park2022self,sun2022contrast}.
Specifically, they only utilize facial videos without supervised labels to train the models.
These methods generate positive ($i.e.,$ similar HRs) and negative ($i.e.,$ dissimilar HRs) pairs by means of frequency resampler~\cite{gideon2021way}, frequency modulation~\cite{yue2022video}, sparsity-based temporal augmentation~\cite{wang2022self,park2022self}, and contrast between two different facial videos~\cite{sun2022contrast}.
Contrastive learning is then applied to pull together the positive pairs and push away the negative pairs.
Contrastive learning, which is commonly used in self-supervised approaches~\cite{van2018representation,he2020momentum,chen2020simple}, is a representation learning method without supervised labels.
Fusion ViViT~\cite{park2022self} is a self-supervised transformer-based method that predicts pulse waves from RGB and near-infrared (NIR) videos.
It performs self-supervised learning using videos captured simultaneously by RGB and NIR cameras, but both RGB and NIR videos are required for inference.
This presents a challenge because it is expensive to provide two different types of cameras for monitoring vital signs.
Contrast-Phys~\cite{sun2022contrast} is a self-supervised 3DCNN-based method that predicts pulse waves from facial videos.
It pushes away pulse waves predicted from two different videos ($i.e.,$ negative pairs) since the HRs of different videos are not similar, while pulse waves predicted from the same video in different spatiotemporal locations ($i.e.,$ positive pairs) are pulled together.
Contrast-Phys randomly samples several small windows from a short 10-second clip as positive pairs.
However, HRs change over ten seconds, especially during exercise, so defining them as positive pairs is not reasonable.
This non-ideal definition could harm model training and degrade the subsequent HR estimation performance.

Our proposed CalibrationPhys considers pulse waves predicted from video clips captured simultaneously by two cameras to be positive pairs.
Thus, since there is no time gap between positive pairs, appropriate model training is possible.
In addition, covering a rich set of negative pairs leads to better representation learning~\cite{he2020momentum,akamatsu2023edema}, but Contrast-Phys has only negative pairs formed from two different clips in the mini-batch.
In contrast, CalibrationPhys can form negative pairs from several different clips in the mini-batch ($e.g.,$ 32 clips), thus leading to better representation learning.
Also, unlike FusionViViT, CalibrationPhys only requires video captured by a single camera for inference.
We build a specific model for each camera to prevent the estimation performance from being degraded by camera differences.
Furthermore, unlike previous self-supervised approaches~\cite{gideon2021way,yue2022video,wang2022self,park2022self,sun2022contrast}, CalibrationPhys also attempts to predict respiratory waves from facial video in a self-supervised manner.

\begin{figure*}[t]
\begin{center}
\includegraphics[scale=0.38]{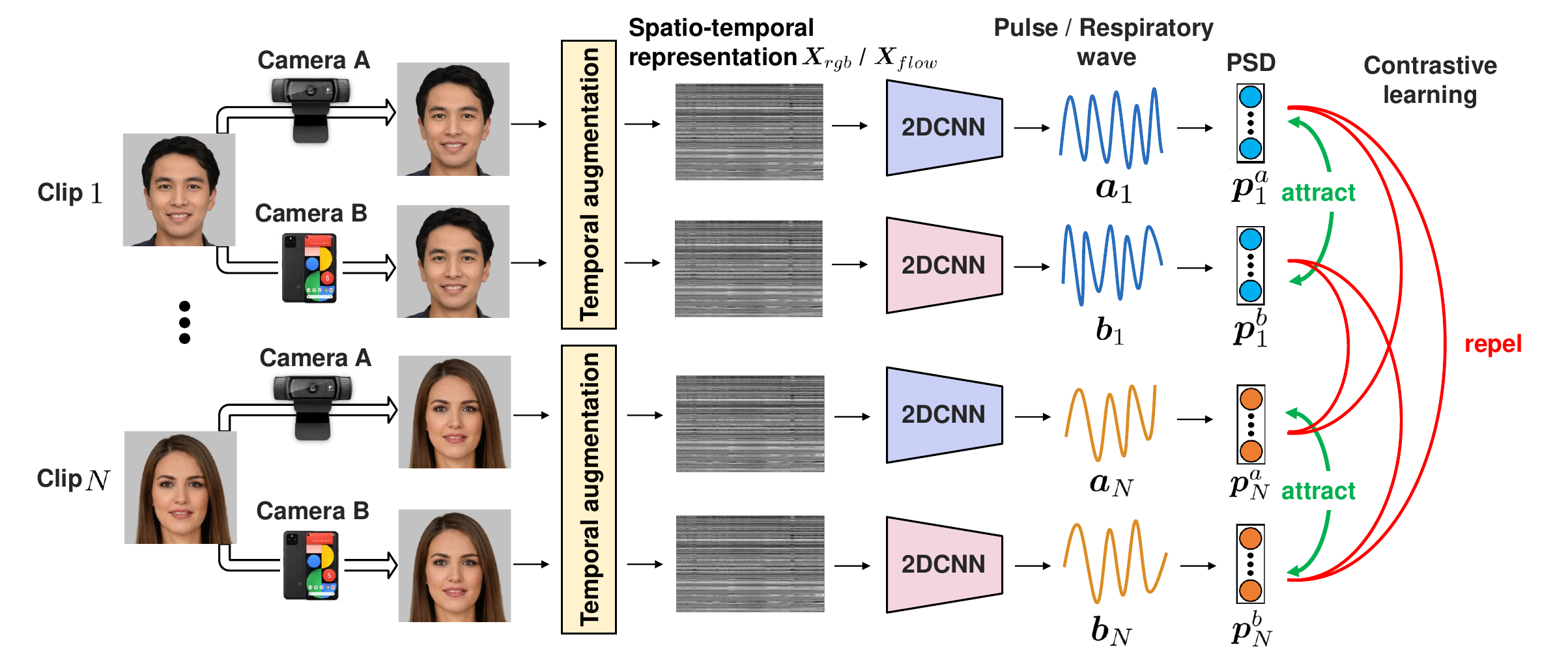}
\end{center}
\caption{Training process of CalibrationPhys. The synchronized videos using multiple cameras are transformed into spatio-temporal representations after temporal augmentation, and the pulse or respiratory waves are predicted via a 2DCNN. We then perform contrastive learning so that HR or RR estimated from facial videos captured simultaneously by two cameras A and B are attracted, and their values estimated from different facial videos are repelled. Note that the models for HR and RR estimation are trained individually.}
\label{fig:method}
\end{figure*}

\section{Method}
\label{sec:method}
All procedures in this study were approved by the NEC Ethical Review Committee for the Life Sciences (approval number: 22-005, date of approval: September 8, 2022).
Informed consent was obtained from all subjects who participated in this study.
Figure~\ref{fig:method} shows the training process of CalibrationPhys.
We present the details of the spatio-temporal representation in \ref{subsec:spatio-temporal}, the temporal augmentation and model network in \ref{subsec:aug}, and the contrastive learning in \ref{subsec:contrastive}.

\begin{figure}[t]
\begin{center}
\includegraphics[scale=0.38]{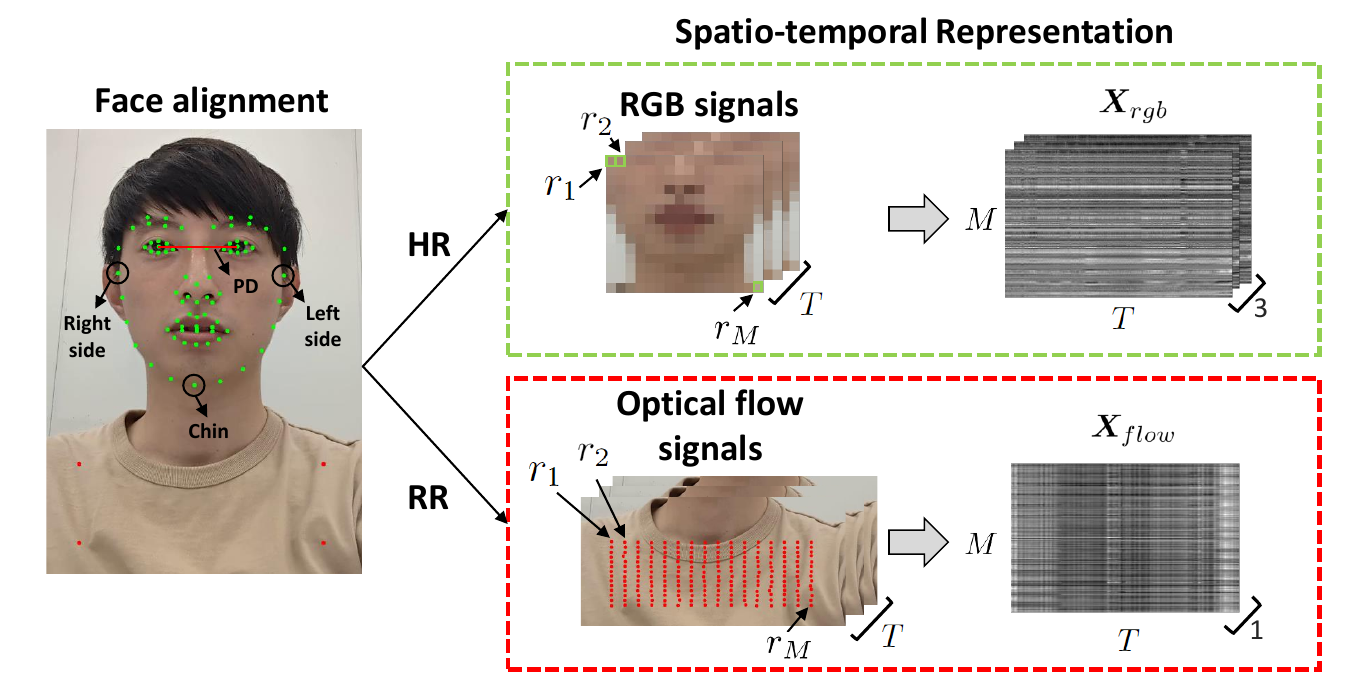}
\end{center}
\caption{Extraction of RGB and optical flow signals from a facial video and construction of spatio-temporal representations.}
\label{fig:input}
\end{figure}

\subsection{Spatio-temporal Representation}
\label{subsec:spatio-temporal}
We utilize facial videos including the face and chest regions, with different input features used for the HR and RR measurements.
For the HR measurement, the RGB signal extracted from the face is utilized to capture subtle changes in reflected light on the skin.
For the RR measurement, the optical flow signal extracted from the chest is utilized to capture chest movements associated with respiratory motion.
Figure~\ref{fig:input} illustrates the extraction of RGB and optical flow signals from a facial video and the construction of spatio-temporal representations.

First, we apply a face alignment method~\cite{imaoka2021future} to a facial video and detect the facial landmarks (indicated by the green points on the left side of Fig.~\ref{fig:input}).
For the HR measurement, we define $16 \times 14$ blocks within the rectangle from below the eyes and generate multiple ROIs by arranging them.
Facial landmarks are detected in every frame and multiple ROIs are generated.
Then, we construct a spatio-temporal representation $\bm{X}_{rgb} \in \mathbb{R}^{3 \times M \times T}$ by using the temporal signals of the three channels (red, blue, and green) in the multiple ROIs.
Here, 3 is the number of channels, $M$ is the number of ROIs including $r_1, \cdots, r_M$, and $T$ is the number of frames ($M = 224$ and $T = 150$ in our experiment).
The proposed method can reduce the computational cost by utilizing 2D spatio-temporal representation instead of 3D representations consisting of 2D spatial and 1D temporal representations used by other methods~\cite{yu2019remote,sun2022contrast}.

For the RR measurement, we identify the chest region (indicated by the four red points on the left in Fig.~\ref{fig:input}) by using the facial landmarks.
Specifically, the horizontal direction of the chest region is in the range $[x_{right}-PD/2,x_{left}+PD/2]$ and the vertical direction is in the range $[y_{chin}+PD,y_{chin}+2*PD]$.
Note that $x_{right}$ and $x_{left}$ are the horizontal coordinates of the right and left sides of the face, $y_{chin}$ is the vertical coordinate of the chin, and $PD$ denotes the pupil distance (see Fig.~\ref{fig:input}).
Next, we extract $16\times14$ optical flow signals from the chest region and generate multiple ROIs by aligning them in the same way as for the HR measurement. 
The chest region is identified only in the initial frame, and the optical flow signal is extracted in subsequent frames.
Optical flow captures the object motion between frames in a video, and in this case, we extract the chest movement associated with respiratory motion. 
Since respiratory motion has major energy in the vertical direction~\cite{zhan2020revisiting}, we use only the vertical optical flow signal.
We then construct a spatio-temporal representation $\bm{X}_{flow} \in \mathbb{R}^{1 \times M \times T}$ by using the optical flow signals in the multiple ROIs. 
Here, 1 is the vertical direction signal based on optical flow, and the number of ROIs and frames is set to $M=224$ and $T=300$, respectively, in our experiment.

\subsection{Temporal Augmentation and Model Network}
\label{subsec:aug}
Since the variation of HR and RR in the training dataset is limited, we introduce temporal augmentation as a data augmentation technique.
First, the number of frames of the spatio-temporal representation is randomly selected from the range of $[T-T_{aug-},T+T_{aug+}]$, and then, the number of frames is transformed to $T$ by upsampling or downsampling the spatio-temporal representation.
Here, $T_{aug-}$ and $T_{aug+}$ represent the frame ranges of temporal augmentation, which we set to 30 and 60, respectively.
For example, by upsampling a 130-frame spatio-temporal representation to a 150-frame one ($i.e.$, slowing down the video), it is feasible to decrease the HR and RR corresponding to the video.
Similarly, downsampling a 170-frame spatio-temporal representation to a 150-frame one ($i.e.$, speeding up the video) could increase the HR and RR.
We perform up/downsampling only in the temporal direction of the spatio-temporal representation by using bilinear interpolation.
Temporal augmentation expands the HR and RR in the training dataset and improves the robustness of the HR and RR measurements.
This idea is similar to the data augmentation technique in a previous HR estimation method~\cite{niu2019robust} that increases or decreases the HR of the supervised labels in accordance with the spatio-temporal representation.
Since we do not use supervised labels, we perform the same up/downsampling on the spatio-temporal representation extracted from the facial videos captured by the two cameras (see Fig.~\ref{fig:method}).

\begin{figure}[t]
\begin{center}
\includegraphics[scale=0.47]{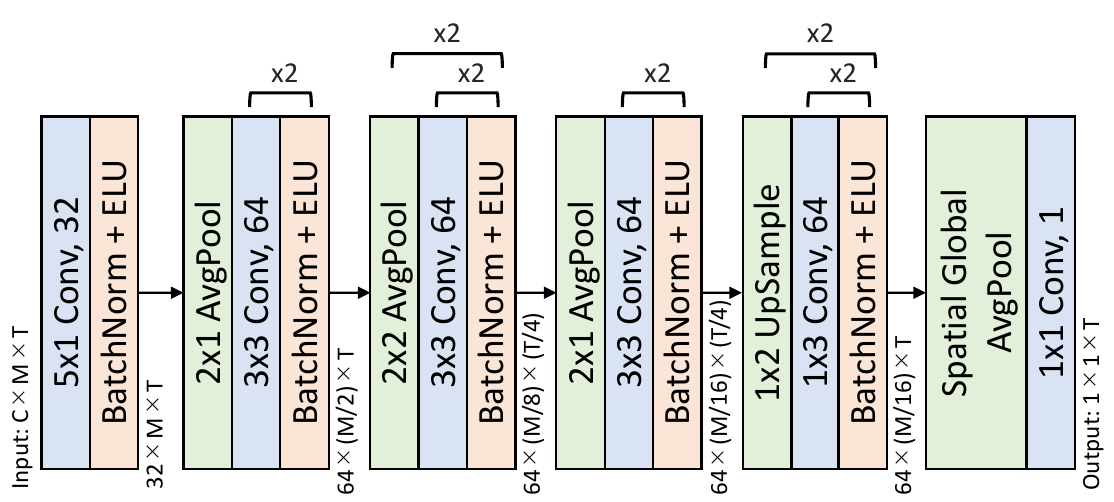}
\end{center}
\caption{Our model network for HR and RR measurements, which is a 2DCNN based on PhysNet~\cite{yu2019remote}. The input is a tensor of $C \times M \times T$ ($C$, $M$, and $T$ are the number of channels, ROIs, and frames, respectively), and the output shape is $1 \times 1 \times T$. ``$5\times1$ Conv, 32" represents a convolution layer with a kernel size of $5\times1$ and 32 channels.}
\label{fig:model}
\end{figure}

We built a PhysNet~\cite{yu2019remote}-based 2DCNN model, and its details are shown in Fig.~\ref{fig:model}. 
State-of-the-art methods for HR measurement~\cite{gideon2021way,sun2022contrast} also utilize PhysNet, which is a model network that outputs pulse waves from facial videos.
The original PhysNet is a 3DCNN that handles a 3D representation consisting of 2D spatial and 1D temporal representations, making it computationally expensive compared to a 2DCNN.
In particular, when we implement it on a device with limited CPU performance and memory (such as a smartphone), it takes a lot of computation time, and the amount of memory may be insufficient.
We have therefore developed a 2DCNN that handles a 2D representation consisting of 1D spatial (by arranging 2D representations in a row) and 1D temporal representations, thus enabling us to reduce the computational cost on smartphones.

\subsection{Contrastive Learning}
\label{subsec:contrastive}
CalibrationPhys is trained by contrastive learning using pulse or respiratory waves produced by the 2DCNN.
In contrastive learning, HR or RR values estimated from facial videos captured simultaneously by two cameras are trained to be similar ($i.e.$, positive pairs), and their values estimated from different facial videos are trained to be dissimilar ($i.e.$, negative pairs).
Note that the models for HR and RR estimation are trained individually.

As shown in Fig.~\ref{fig:method}, suppose that $\bm{A}=[\bm{a}_1,\cdots,\bm{a}_N]$ and $\bm{B}=[\bm{b}_1,\cdots,\bm{b}_N]$ are pulse or respiratory waves produced by the 2DCNN using the videos captured by camera A and B, respectively.
Here, $\bm{a}_n$ and $\bm{b}_n$ are vectors representing the pulse or respiratory waves in the $n$th video clip, and $N$ is the number of video clips in the mini-batch.
Then, the power spectrum densities (PSDs) $\bm{P}_A=[\bm{p}_1^a,\cdots,\bm{p}_N^a]$ and $\bm{P}_B=[\bm{p}_1^b,\cdots,\bm{p}_N^b]$ are calculated from $\bm{A}$ and $\bm{B}$, respectively.
The PSD represents frequency characteristics, and we can obtain HR and RR information from the pulse and respiratory waves.
To obtain frequency bands related to HR and RR, we calculated PSDs with respect to 0.667--2.50 Hz ($i.e.$, HR at 40--150 beats per minute (bpm)) and 0.167--1.0 Hz ($i.e.$, RR at 10--60 breaths per minute (bpm)), respectively.
In CalibrationPhys, we design the following loss function $\mathcal{L}$ based on InfoNCE loss~\cite{van2018representation}:
\begin{flalign}
    &\mathcal{L} = \sum_{n=1}^{N} (\mathcal{L}_n^a + \mathcal{L}_n^b),
\end{flalign}
where
\begin{flalign}
    & \mathcal{L}_n^a =  \log \frac{\exp(d(\bm{p}_n^a,\bm{p}_n^b) /\tau)}{\sum_{k=1}^{N} (\1_{k \neq n} \exp(d(\bm{p}_n^a,\bm{p}_k^a)/\tau)+\exp(d(\bm{p}_n^a,\bm{p}_k^b)/\tau))}, \nonumber \\ 
    & \mathcal{L}_n^b = 
    \log \frac{\exp(d(\bm{p}_n^b,\bm{p}_n^a) /\tau)}{\sum_{k=1}^{N} (\1_{k \neq n} \exp(d(\bm{p}_n^b,\bm{p}_k^b)/\tau)+\exp(d(\bm{p}_n^b,\bm{p}_k^a)/\tau))}. \nonumber
\end{flalign}
In Eq. (1), $\mathcal{L}_n^a$ and $\mathcal{L}_n^b$ are the loss functions for cameras A and B, respectively, $d$ is the mean squared error, $\tau$ is the temperature parameter, and $\1_{k \neq n} \in \{0,1\}$ is a function that returns 1 when $k \neq n$ and 0 otherwise.
The loss function $\mathcal{L}$ forces the HR or RR estimated from facial videos captured simultaneously by two cameras to be attracted, and those values estimated from different facial videos to be repelled.
In addition, while Contrast-Phys~\cite{sun2022contrast} uses two different pairs of facial video clips within a mini-batch for contrastive learning, CalibrationPhys can use $N$ facial video clips.
Since more negative pairs can be used, CalibrationPhys can acquire better feature representations compared to Contrast-Phys.

\section{Experiments}
\label{sec:experiment}
We present the dataset in \ref{subsec:dataset}, the experimental settings in \ref{subsec:setting}, and the results in \ref{subsec:result}.

\subsection{Datasets}
\label{subsec:dataset}
\subsubsection{Multi-camera Dataset}
Self-collected dataset includes 20 subjects using multiple cameras.
The 20 subjects consisted of 11 males and 9 females, one in his/her 20s, 4 in their 30s, 9 in their 40s, and 6 in their 50s.
The subjects are mainly Japanese, and their skin color type is Asian race.
We collected facial videos, pulse waves, HRs, and respiratory waves from the subjects.
Facial videos were captured simultaneously using two devices: a Logitech C920n HD PRO webcam and a Google Pixel~5 smartphone.
We selected one webcam and one smartphone for actual usage scenarios.
Facial videos were captured at a resolution of $1920\times1080$, a frame rate of 30 fps, and at a distance of 50 cm from the subject.
Pulse waves were acquired using the wearable sensor E4 wristband and utilized as supervised labels to train DL models (not for our method but for the comparative methods).
HRs were obtained by the OxyTrue A pulse oximeter~\footnote[6]{\url{https://www.medacx.co.uk/products/pulse-oximeters/oxytrue-handheld-pulse-oximeter}} and used as the ground-truth for evaluating the HR estimation performance.
Respiratory waves were acquired by the RIP respiratory sensor and used as supervised labels (for comparative methods) and as the ground-truth for evaluating the RR estimation performance.
The respiratory sensor was attached to the chest and respiratory waves were acquired. 
The RIP respiratory sensor measures the overall displacement of the thorax or abdomen, making it more immune to motion-induced artifacts \protect \footnotemark[4].

For each subject, facial videos were captured in two sessions, one with the smartphone placed on a stand and the other with it held in the hand.
The webcam was fixed to the stand in both sessions.
In each session, four tasks were performed for 60 seconds each: resting (T1), breath holding (T2), foot stomping (T3), and talking (T4) to vary the HR and RR.
In T2 and T3, breath holding or foot stomping was performed for approximately 30 seconds, and the remaining time was for resting.
In this experiment, subjects were seated to test the possibility of using the technology in online meetings and telemedicine.
Breath holding and foot stomping were performed to rapidly change HR and RR, and conversations were validated for use in online meetings.
We considered 60 seconds in each task as a single video, and videos with inappropriate pulse or respiratory waves were excluded.
Table~\ref{table:condition} summarizes these experimental tasks.

\begin{table}[t]
\caption{Experimental tasks}
\label{table:condition}
\vspace{-10pt}
\begin{center}
\scalebox{0.9}{
\begin{tabular}{c|l}  \hline
Session 1 & Smartphone placed on a stand \\
Session 2 & Smartphone held in the hand \\ \hline
T1 & Rest [60s] \\ 
T2 & Breath holding [30s] + Rest [30s] \\ 
T3 & Foot stomping [30s] + Rest [30s] \\
T4 & Talking [60s] \\ \hline
\end{tabular}
}
\end{center}
\end{table}

\subsubsection{Pre-train Dataset}
Self-collected dataset for pre-training includes 22 subjects using a webcam.
The 22 subjects consisted of 9 males and 13 females, 2 in their 20s, 5 in their 30s, 9 in their 40s, and 6 in their 50s.
The subjects are mainly Japanese, and their skin color type is Asian race.
This dataset was used for CalibrationPhys w/ Pre-train.
We collected facial videos, pulse waves, HRs, and respiratory waves from the subjects.
Facial videos were captured by a Logitech C920n HD PRO webcam.
Therefore, we constructed a pre-trained model for the webcam (corresponding to Camera A in Fig.~\ref{fig:intro}~(ii)), which was trained using supervised labels.
Sensor devices for the pulse waves, HRs, and respiratory waves were the same as those in the multi-camera dataset, as were the experimental tasks.

\subsubsection{UBFC-rPPG~\cite{bobbia2019unsupervised}}
Public dataset includes 42 subjects using a webcam.
Facial videos and pulse waves were collected from the subjects.
Facial videos were captured by the same Logitech C920n HD PRO webcam as our self-collected dataset.
The length of each video was about 60 seconds.
This dataset was used to evaluate the HR estimation performance of the webcam model trained with the multi-camera dataset ($i.e.$, cross-dataset evaluation).
CalibrationPhys constructs camera-specific models.
Since the muti-camera and UBFC-rPPG datasets use the same webcam, we used the UBFC-rPPG dataset for cross-dataset evaluation.

\subsection{Experimental Settings}
\label{subsec:setting}
Using the multi-camera dataset, we performed a four-fold cross-validation in which 15 subjects were used as training data and five as test data.
In other words, we evaluated the HR and RR estimation performance for new subjects not included in the training data.
Among all videos for training data, $20\%$ were used as validation data.
From each 60-second video in the training data, we randomly selected short video clips including $T$ frames for the HR ($T=150$) and RR ($T=300$) measurements.
The values of $T$ were determined experimentally.
The estimation performance corresponding to other values of $T$ is shown in Appendix~A.
The clips were selected from the webcam and smartphone video at the same initial frame and then utilized to train CalibrationPhys.
For CalibrationPhys w/ Pre-train, pre-trained models for the HR and RR measurements were built using facial videos captured by the webcam, pulse waves, and respiratory waves from the pre-train dataset.
The pre-trained models were optimized based on Eq.~(1) (which is the same loss function as CalibrationPhys) and then trained based on the output waves from the webcam and the ground-truth data measured by a pulse wave sensor or respiratory belt.
For all models, we trained for 30 epochs and used the model at the epoch with
the lowest validation loss for testing.
The optimization algorithm was AdamW~\cite{loshchilov2018decoupled}, the batch size was 32, and the learning rate was 1e-3 for the HR measurement and 1e-4 for the RR measurement.
The temperature parameter $\tau$ was set to 0.1 according to refs.~\cite{khosla2020supervised,akamatsu2023edema}.
The framework was implemented with PyTorch~\cite{paszke2019pytorch} on eight NVIDIA A100 GPUs.
When the minibatch size was multiplied by $k$ due to the multi-GPU, the above learning rate was multiplied by $k$ according to the linear scaling rule~\cite{goyal2017accurate}.

For the inference, pulse and respiratory waves were estimated from a 60-second video in each task.
We then applied a fast Fourier transform (FFT)~\footnote[7]{Frequency range is 0.75--2.50 Hz (45--150 bpm) for HR measurement and 0.167--0.667 Hz (10--40 bpm) for RR measurement.} to the waves in a 10-second window size with a 1-second step and calculated the HR and RR for 60 seconds (one point per second).
In contrast to most previous studies, which have calculated HR and RR with longer window sizes ($e.g.$, 30 seconds~\cite{sun2022contrast}), we use a shorter window size of ten seconds.
Our motivation is that the measurement results should be updated frequently for users.
While this enables HR and RR to be calculated in a shorter time, it is a more challenging task because we have less information to calculate HR and RR.

\subsection{Results}
\label{subsec:result}

\subsubsection{Differences in HR and RR estimation performance between webcam and smartphone}
\begin{figure}[t]
\begin{center}
\includegraphics[scale=0.57]{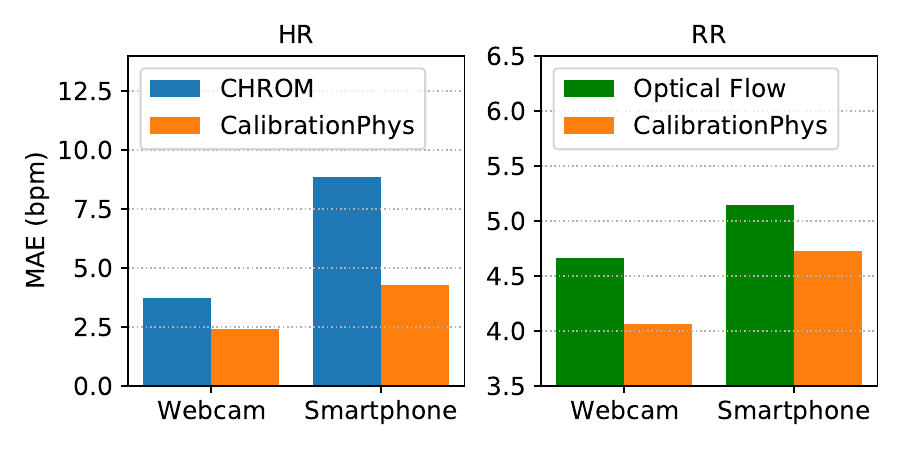}
\end{center}
\vspace{-10pt}
\caption{Mean absolute error (MAE) of HR (left) and RR (right) measurements between webcam and smartphone in the multi-camera dataset. In the HR measurement, CHROM~\cite{de2013robust} and CalibrationPhys are compared. In the RR measurement, Optical Flow and CalibrationPhys are compared.}
\label{fig:mae}
\end{figure}

Figure~\ref{fig:mae} shows the mean absolute error (MAE) of the HR and RR measurements between the webcam and smartphone in the multi-camera dataset.
In the HR measurement, we compare our CalibrationPhys with CHROM~\cite{de2013robust}, a popular signal processing HR estimation method.
In the RR measurement, we compare our CalibrationPhys with optical flow, a benchmark RR estimation method.
All methods are built without supervised labels.

First, the results of CHROM indicate that the HR estimation performance on the smartphone is worse than that on the webcam.
This is possibly because each camera has different colors, internal image processing, camera settings ($e.g.$, white balance and exposure), and capturing conditions ($e.g.$, on a stand and hand-held).
Therefore, the adoption of one algorithm for both cameras has limitations in terms of estimation performance.
Second, CalibrationPhys can improve the HR estimation performance for both cameras, especially for the smartphone.
CalibrationPhys improves the HR estimation performance by training the specific DL model for each camera, and its strength is that supervised labels are not required for the training.

The results of optical flow in the RR measurement also show that the performance on the smartphone is worse than that on the webcam.
This is because the estimation performance is degraded due to camera shake in the RR measurement on the smartphone (see \ref{subsec:rr-result} hereafter).
CalibrationPhys has a better RR estimation performance than optical flow for both cameras.

\subsubsection{HR estimation performance on smartphone}
Since HR measurement on the smartphone is a more challenging task than on the webcam (Fig.~\ref{fig:mae}), we compare CalibrationPhys with state-of-the-art methods for HR measurement on the smartphone.
As comparative methods, we used three signal processing methods~\footnote[8]{Signal processing methods were implemented by using pyVHR~\cite{boccignone2020open,boccignone2022pyvhr}.} (ICA~\cite{poh2010non}, CHROM~\cite{de2013robust}, and POS~\cite{wang2016algorithmic}), four supervised DL-based methods~\footnote[9]{DL-based methods were implemented by using rPPG-Toolbox~\cite{liu2023rppgtoolbox}.} (DeepPhys~\cite{chen2018deepphys}, PhysNet~\cite{yu2019remote}, TS-CAN~\cite{liu2020multi}, and EfficientPhys-C~\cite{liu2023efficientphys}), and one self-supervised DL-based method (Contrast-Phys~\cite{sun2022contrast}).
EfficientPhys-C~\cite{liu2023efficientphys} and Contrast-Phys~\cite{sun2022contrast} are the state-of-the-art supervised and self-supervised methods, respectively, that have been proposed within the last two years.
For the signal processing methods, a single ROI was utilized as input by averaging the RGB signals for the same ROI as the proposed method.
For the DL-based methods, multiple ROIs sized $36\times36$ were generated for the same ROI as the proposed method and input to the models.
The other experimental settings were the same as the proposed method.
We also calculated the computational cost of the DL-based methods.
Specifically, we examined the number of model parameters and the computational complexity by FLOPs when inputting a 10-second facial video.

Table~\ref{table:hr} shows the MAE and computational cost for the HR estimation on the smartphone.
As we can see, CalibrationPhys achieves a better HR estimation performance than the state-of-the-art methods.
First, CalibrationPhys outperforms all signal processing methods, which indicates that deep neural networks can significantly improve the HR estimation performance.
Second, CalibrationPhys has a better performance than the supervised DL-based methods.
It is very beneficial that our method, which does not require supervised labels, achieves a better performance than supervised methods.
Third, CalibrationPhys also outperforms Contrast-Phys, which is a state-of-the-art self-supervised DL-based method.
Our method is particularly effective on the two more challenging tasks T3 (foot stomping) and T4 (talking).
Contrast-Phys may incorrectly create positive pairs in tasks with large HR changes, resulting in the poor HR estimation performance in challenging tasks.
CalibrationPhys performs contrastive learning based on two synchronized videos, making its training reasonable even with rapid HR changes.
This is presumably why our method achieves robust HR measurements even for T3 and T4.
Furthermore, CalibrationPhys is superior to other methods in terms of computational cost.
Specifically, the number of parameters is more than twice as small compared to Contrast-Phys, and the FLOPs are nearly ten times smaller.

\begin{table*}[t]
\caption{Mean absolute error (MAE) and computational cost for HR estimation on the smartphone in the multi-camera dataset. The MAE for each task and the MAE for all tasks in the session are shown.}
\label{table:hr}
\begin{center}
\scalebox{0.9}{
\begin{tabular}{l|ccccc|ccccc|cc}  \hline
\multicolumn{1}{c|}{} & \multicolumn{5}{c|}{Session 1} & \multicolumn{5}{c|}{Session 2} & \multicolumn{2}{c}{Computational cost}
\\ 
\cline{2-13}
Method &T1 & T2 & T3 & T4 & All & T1 & T2 & T3 & T4 & All & \#Param. & FLOPs  \\ \hline\hline
\textit{Signal processing}: & & & & & & & & & & &  \\
ICA~\cite{poh2010non} & 11.15 & 11.66 & 15.98 & 16.10 & 13.52 & 11.33 & 11.04 & 12.54 & 14.96 & 12.33 & -- & --  \\ 
CHROM~\cite{de2013robust} & 6.15 & 7.50 & 10.52 & 12.29 & 8.93 & 7.45 & 6.54 & 8.67 & 13.52 & 8.81 & -- & -- \\ 
POS~\cite{wang2016algorithmic} & 7.48 & 8.99 & 12.62 & 14.40 & 10.66 & 7.75 & 7.90 & 10.20 & 16.09 & 10.19 & -- & --  \\  \hline
\textit{Supervised DL}: & & & & & & & & & & &  \\
DeepPhys~\cite{chen2018deepphys} & 3.44 & 3.92 & 11.30 & 12.15 & 7.37 & 4.75 & 5.51 & 9.88 & 13.00 & 7.97 & 0.533M & 15.47G \\
PhysNet~\cite{yu2019remote} & 3.19 & 3.89 & 7.60 & 8.29 & 5.56 & 3.79 & 4.57 & 5.48 & 9.16 & 5.58 & 0.736M & 10.29G \\
TS-CAN~\cite{liu2020multi} & 2.87 & 3.95 & 8.30 & 11.28 & 6.35 & 3.55 & 4.61 & 7.85 & 10.17 & 6.30 & 0.533M & 15.47G \\
EfficientPhys-C~\cite{liu2023efficientphys} & 2.87 & 3.99 & 8.46 & 8.67 & 5.78 & 3.17 & {\bf 3.75} & 6.39 & 10.13 & 5.61 & 0.467M & 7.806G \\ 
\hline
\textit{Self-supervised DL}: & & & & & & & & & & &  \\
Contrast-Phys~\cite{sun2022contrast} & 3.21 & 3.89 & 9.32 & 9.90 & 6.33 & 3.55 & 4.45 & 8.21 & 9.78 & 6.26 & 0.858M & 20.23G  \\
\bf{CalibrationPhys}  & {\bf 2.76} & {\bf 3.58} & {\bf 4.10} & {\bf 5.81} & {\bf 4.00} & {\bf 2.80} & 4.26 & {\bf 4.58} & {\bf 7.03} & {\bf 4.55} & {\bf 0.304M} & {\bf 2.858G}   \\\hline
\end{tabular}
}
\end{center}
\end{table*}

\begin{table*}[t]
\caption{Mean absolute error (MAE) and computational cost for RR estimation on the smartphone in the multi-camera dataset. The MAE for each task and the MAE for all tasks in the session are shown.}
\label{table:rr}
\begin{center}
\scalebox{0.9}{
\begin{tabular}{l|ccccc|ccccc|cc}  \hline
\multicolumn{1}{c|}{} & \multicolumn{5}{c|}{Session 1} & \multicolumn{5}{c|}{Session 2} & \multicolumn{2}{c}{Computational cost}
\\ 
\cline{2-13}
Method & T1 & T2 & T3 & T4 & All & T1 & T2 & T3 & T4 & All & \#Param. & FLOPs  \\ \hline\hline
\textit{Benchmark}: & & & & & & & & & & & \\
RGB & 3.83 & 5.68 & 6.64 & 5.79 & 5.49 & 6.23 & 6.12 & 8.25 & 5.76 & 6.61 & -- & --    \\ 
Optical Flow & 1.56 & 5.08 & 4.66 & 4.48 & 3.95 & 5.83 & 6.12 & 8.02 & {\bf 5.28} & 6.34 & -- & -- \\ \hline
\textit{Supervised DL}: & & & & & & & & & & & \\
DeepPhys~\cite{chen2018deepphys} & 4.85 & 5.83 & 7.55 & 5.20 & 5.86 & 6.61 & 5.79 & 8.61 & 5.55 & 6.65 & 0.533M & 15.47G  \\
TS-CAN~\cite{liu2020multi} & 5.09 & 6.00 & 7.58 & 5.25 & 5.98 & 5.93 & 5.91 & 8.20 & 5.43 & 6.39 & 0.533M & 15.47G \\ \hline
\textit{Self-supervised DL}: & & & & & & & & & & & \\
\bf{CalibrationPhys} & {\bf 1.55} & {\bf 4.90} & {\bf 3.57} & {\bf 4.46} & {\bf 3.62} & {\bf 5.18} & {\bf 5.75} & {\bf 6.98} & 5.32 & {\bf 5.83} & {\bf 0.303M} & {\bf 2.836G}  \\\hline
\end{tabular}
}
\end{center}
\end{table*}

\subsubsection{RR estimation performance on smartphone}
\label{subsec:rr-result}
Since RR estimation on the smartphone is also a more challenging task than on the webcam (Fig.~\ref{fig:mae}), we compare CalibrationPhys with state-of-the-art methods for RR estimation on the smartphone.
As comparative methods, we used two benchmark methods (RGB and Optical Flow) and two supervised DL-based methods (DeepPhys~\cite{chen2018deepphys} and TS-CAN~\cite{liu2020multi}).
The benchmark using RGB signals corresponds to pixel intensity variation-based RR estimation, where a single ROI was used as input by averaging RGB signals into a single channel for the same ROI as the proposed method.
Another benchmark using optical flow signals corresponds to pixel movement-based RR measurement, where a single ROI was used as input by averaging optical flow signals for the same ROI as the proposed method.
The DL-based methods take RGB signals as inputs, where multiple ROIs sized $36\times36$ were generated for the facial ROI, which was used in the original studies~\cite{chen2018deepphys,liu2020multi}.
The other experimental settings were the same as the proposed method.

Table~\ref{table:rr} shows the MAE and computational cost for the RR estimation on the smartphone.
As we can see, CalibrationPhys is again superior to the other methods.
As a previous study~\cite{zhan2020revisiting} reported, the results suggest that pixel movement-based methods ($i.e.$, Optical Flow and CalibrationPhys) have better RR estimation performance than pixel intensity variation-based methods ($i.e.$, RGB, DeepPhys, and TS-CAN). 
For all methods, RR estimation performance when holding the smartphone in the hand (Session 2) is lower than when placing the smartphone on the stand (Session 1).
This is because the camera shake when holding the smartphone results in a failure to successfully capture the chest movement.
Nevertheless, CalibrationPhys achieves a more accurate RR estimation than Optical Flow by training a model that is robust to noise.
In this case, we can suppress the noise by means of contrastive learning between the smartphone and the webcam fixed to the stand, which is not affected by camera shake.
The computational cost of CalibrationPhys is also lower than the other DL-based RR measurement methods.
For HR and RR measurements, CalibrationPhys uses PhysNet-based 2DCNN (see~\ref{subsec:aug}) as the backbone.
Performance comparisons when using commonly used CNNs as the backbone are discussed in Appendix~B.
Furthermore, performance comparisons of HR and RR on the webcam with state-of-the-art methods are discussed in Appendix C.

\begin{table}[t]
     \caption{MAE for HR and RR estimation on the smartphone in the multi-camera dataset.}
    \begin{subtable}[h]{0.45\textwidth}
        \centering
        \caption{HR}
        \scalebox{0.9}{
        \begin{tabular}{l|ccc}  \hline
        Method &  Session 1 & Session 2 & Mean \\ \hline
        CalibrationPhys & 4.00 & 4.55 & 4.28  \\
        {\bf CalibrationPhys w/ Pre-train} & {\bf 3.83} & {\bf 4.35} & {\bf 4.09}    \\
        \hline
        \end{tabular}
        }
    \end{subtable}
    \hfill
    \vspace{10pt}
    \begin{subtable}[h]{0.45\textwidth}
        \centering
        \caption{RR}
        \scalebox{0.9}{
        \begin{tabular}{l|ccc}  \hline
        Method &  Session 1 & Session 2 & Mean \\ \hline
        CalibrationPhys & 3.62 & 5.83 & 4.73  \\
        {\bf CalibrationPhys w/ Pre-train} & {\bf 3.54} & {\bf 5.75} & {\bf 4.65}    \\
        \hline
        \end{tabular}
        }
     \end{subtable}
     \label{table:pre-mae}
\end{table}

\subsubsection{CalibrationPhys vs. CalibrationPhys w/ Pre-train}
We compare CalibrationPhys and CalibrationPhys w/ Pre-train.
In CalibtarionPhys w/ Pre-train, we perform contrastive learning between a model for the smartphone~\footnote[10]{Although the pre-trained model for the webcam could also be used as the model for the smartphone, we experimentally found that it is better to use the pre-trained model for RR estimation and not to use it for HR estimation, and therefore used it in that manner.} and a model for the webcam that has been pre-trained with supervised labels  (see Fig.~\ref{fig:intro}~(ii)).
Table~\ref{table:pre-mae} shows a comparison of CalibrationPhys and CalibrationPhys w/ Pre-train for the MAE of HR and RR estimation on the smartphone in the multi-camera dataset.
The MAE for all tasks in each session and their means are shown.
As we can see, CalibrationPhys w/ Pre-train is superior to CalibrationPhys for both HR and RR estimation.
We investigated the statistical significance of the results with the Wilcoxon signed-rank test using MAEs corresponding to HR and RR estimation in all videos.
The alternative hypothesis is a one-sided test that the MAEs of CalibrationPhys w/ Pre-train are smaller than those of CalibrationPhys.
Its p-value is $0.0358$, and the significance is confirmed ($p<0.05$).
In addition, the pre-training process took only 323 seconds for HR and 249 seconds for RR in our computational environment.
We confirm that the utilization of the pre-trained model can boost HR and RR estimation performance.
This result indicates that if one pre-trained model is used in contrastive learning, the other model also benefits from the pre-training.
Since we consider cases where a pre-trained model for a particular camera is available, it is useful to leverage this pre-trained model.
The result that the other model benefits from the pre-training model in the CalibrationPhys framework is interesting. While not the most important argument of this paper, it is one of the cases that should be considered.

\begin{figure}[t]
\begin{center}
\includegraphics[scale=0.55]{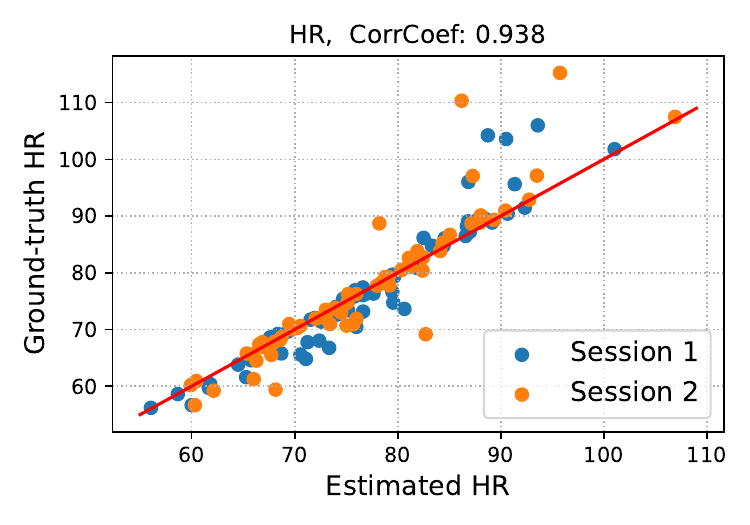}
\includegraphics[scale=0.55]{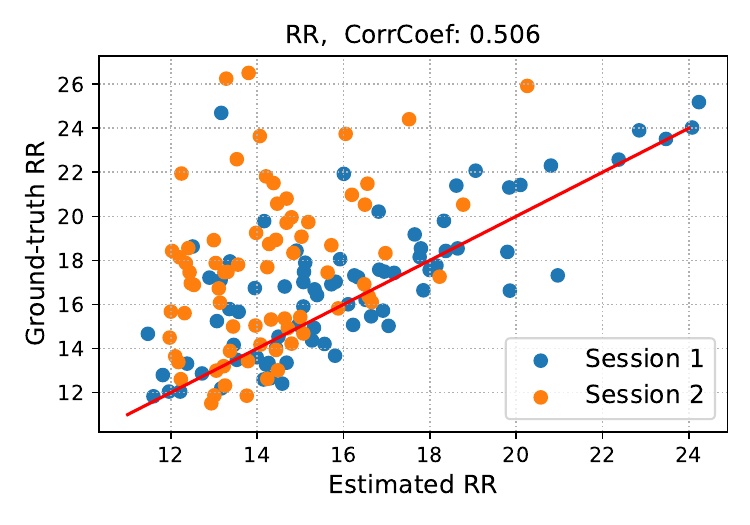}
\end{center}
\vspace{-10pt}
\caption{Correlation plots of estimated and ground-truth results for HR (top) and RR (bottom) measurements on the smartphone in the multi-camera dataset. The colors of the plots are separated for each session, and correlation coefficients (CorrCoef) are displayed in the titles.}
\label{fig:corr}
\end{figure}

Figure~\ref{fig:corr} shows correlation plots of the estimated and ground-truth HR and RR in CalibrationPhys w/ Pre-train.
Note that HR and RR are averages of those values in each 60-second video.
The Pearson correlation coefficient is 0.938 for the HR measurement and 0.506 for the RR measurement.
Several HR estimation cases have poor performance when the ground-truth HR is near 100, which corresponds to challenging tasks such as T3 and T4.
The correlation coefficients for RR estimation in Sessions 1 and 2 are 0.732 and 0.373, respectively.
In Session 2, subtle respiratory movements of the chest cannot be captured due to camera shake (especially in T3), thus causing performance degradation.
Also, since the range of RR is not as wide as that of HR, the correlation coefficient of RR is lower than that of HR.

\subsubsection{Cross-dataset evaluation for HR estimation}
To verify the generalization ability of CalibrationPhys, we perform a cross-dataset evaluation using the multi-camera dataset and UBFC-rPPG~\cite{bobbia2019unsupervised}, a well-known public HR measurement dataset.
Specifically, we train CalibrationPhys and the comparative models on all videos from the multi-camera dataset and directly test them on UBFC-rPPG.
Since the same webcam is used in both datasets, we test the model for the webcam trained on the multi-camera dataset.
The two datasets differ in terms of the subjects' race and HR ranges ($i.e.$, mainly 60--90 bpm in the multi-camera dataset and mainly 80--130 bpm in UBFC-rPPG).
To suppress the gap between these HR ranges, we set the temporal augmentation hyperparameters in CalibrationPhys to $T_{aug-}=0$ and $T_{aug+}=100$.
As in a previous study~\cite{sun2022contrast}, we use MAE, root mean squared error (RMSE), and Pearson correlation coefficient (CorrCoef) as evaluation metrics.
We also use the coefficient of determination $R^2$ for evaluation.

Table~\ref{table:ubfc} lists the results of the cross-dataset evaluation for HR estimation.
As we can see, CalibrationPhys still outperforms the other methods.
These results show that CalibrationPhys trained on a dataset without supervised labels generalizes well to the new dataset.
Thus, we confirm that CalibrationPhys has a high generalization ability.

\begin{table}[t]
\caption{Cross-dataset evaluation for HR estimation on UBFC-rPPG.}
\label{table:ubfc}
\begin{center}
\scalebox{0.9}{
\begin{tabular}{l|cccc}  \hline
Method &MAE$\downarrow$ & RMSE$\downarrow$ & CorrCoef$\uparrow$ & $R^2 \uparrow$ \\ \hline\hline
\textit{Signal processing}: & & &   \\
ICA~\cite{poh2010non} & 19.33 & 24.53 & 0.216 & -1.486   \\ 
CHROM~\cite{de2013robust} & 3.65 & 7.48 & 0.639 & 0.320  \\ 
POS~\cite{wang2016algorithmic} & 3.73 & 7.27 & 0.692 & 0.256  \\  \hline
\textit{Supervised DL}: & & &   \\
DeepPhys~\cite{chen2018deepphys} & 8.26 & 12.64 & 0.505 & -0.135 \\
PhysNet~\cite{yu2019remote} & 3.38 & 7.04 & 0.626 & 0.339  \\
TS-CAN~\cite{liu2020multi} & 3.03 & 6.92 & 0.652 & 0.430  \\
EfficientPhys-C~\cite{liu2023efficientphys} & 3.34 & 6.91 & 0.688 & 0.436  \\ 
\hline
\textit{Self-supervised DL}: & & & \\
Contrast-Phys~\cite{sun2022contrast} & 2.49 & 5.11 & 0.683 & 0.331   \\ 
\bf{CalibrationPhys}  & {\bf 1.83} & {\bf 3.77} & {\bf 0.809} & {\bf 0.627} \\\hline
\end{tabular}
}
\end{center}
\vspace{-5pt}
\end{table}

\section{Discussion}
\label{sec:discussion}
We have demonstrated that CalibrationPhys outperforms state-of-the-art methods in HR and RR measurements.
In this section, we further validate CalibrationPhys, clarify its strengths, and discuss its limitations.

\begin{figure}[t]
\begin{center}
\includegraphics[scale=0.55]{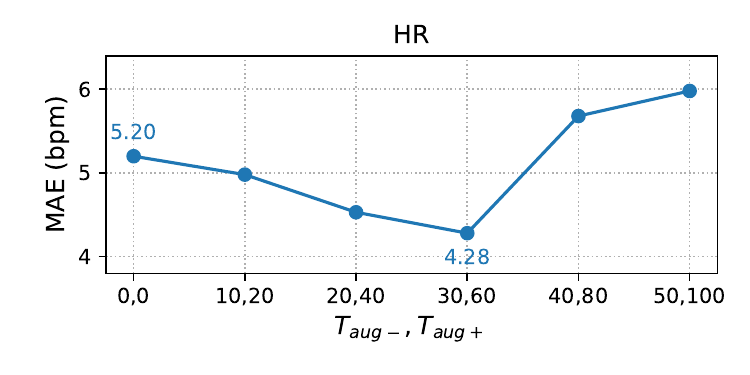}
\includegraphics[scale=0.55]{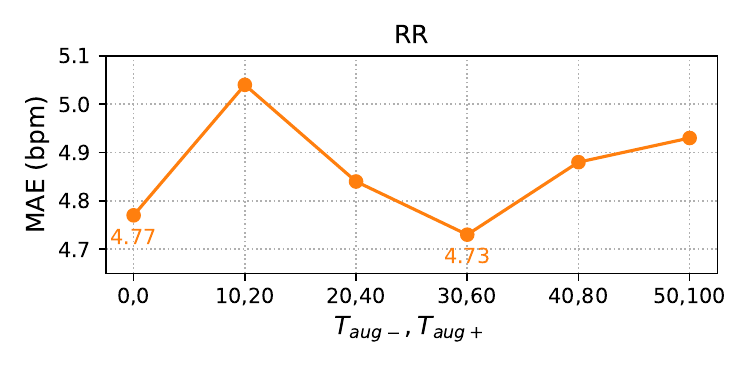}
\end{center}
\vspace{-10pt}
\caption{Relationship between the frame ranges of temporal augmentations $T_{aug-}$, $T_{aug+}$ and MAE for HR and RR estimation on the smartphone in the multi-camera dataset.}
\label{fig:aug}
\end{figure}

One of the key factors contributing to the strong performance of CalibrationPhys is the introduction of temporal augmentation.
Figure~\ref{fig:aug} shows the relationship between the frame ranges of temporal augmentations $T_{aug-}$, $T_{aug+}$ and MAE for HR and RR estimation on the smartphone in the multi-camera dataset.
The means of MAE in Sessions 1 and 2 are shown.
$T_{aug-}=0$ and $T_{aug+}=0$ refer to the settings without temporal augmentation.
For both HR and RR estimation, we find that the estimation performance is improved when $T_{aug-}=30$ and $T_{aug+}=60$ compared to the case without temporal augmentation.
Especially in HR estimation, MAE is reduced by nearly 1 bpm, indicating the significant effectiveness of temporal augmentation.
We observed that the effect is most significant with a 6.41-bpm reduction in MAE when the HR range is high (more than 90 bpm).
Since the range of HRs is wide and it is difficult to cover them all with training data (high HR range of more than 90 bpm in this case), the data augmentation technique is highly effective.

CalibrationPhys uses specific models for two cameras: one for the smartphone and one for the webcam.
We verify the effectiveness of this configuration by testing the models trained for the smartphone and the webcam on both cameras.
We also test a single general model for the smartphone and the webcam, $i.e.$, where the model parameters for the two cameras are shared.
The general model is optimized by weight-sharing with a Siamese network structure~\cite{li2022survey} in CalibrationPhys.
The loss function for the general model is Eq. (1), but the model parameters for cameras A and B are shared.
We verify the effectiveness of constructing camera-specific models by comparing them to the general model.
Table~\ref{table:cross-camera} shows the MAE for HR and RR estimation across different combinations of models and test cameras in the multi-camera dataset.
The MAE for all tasks in each session and their means are shown.
As we can see, providing a specific model for each camera ($e.g.$, when the models for the smartphone are tested on the smartphone) is effective overall.
Crossing models and cameras tends to degrade the estimation performance, especially when models for the webcam are tested on the smartphone.
Furthermore, the general model for the two cameras has a lower HR and RR estimation performance than each specific model.
Even so, a general model for multiple cameras can still be useful in some cases, since only a single model is required for multiple cameras.
Ultimately, we conclude that the specific models for each camera perform better.

\begin{table}[t]
     \caption{MAE for HR and RR estimation across different combinations of models and test cameras in the multi-camera dataset.}
    \begin{subtable}[h]{0.48\textwidth}
        \centering
        \caption{HR}
            \scalebox{0.9}{
            \begin{tabular}{cc|ccc}  \hline
            Model & Test camera & Session 1 & Session 2 & Mean \\ \hline
            Smartphone & Smartphone & 4.00 & {\bf 4.55} & {\bf 4.28}  \\
            Webcam & Smartphone & 5.22 & 6.02 & 5.62    \\
            General & Smartphone & {\bf 3.95} & 4.94 & 4.45    \\ 
            \hline
            Webcam & Webcam & {\bf 2.28} & 2.53 & {\bf 2.41}   \\
            Smartphone & Webcam & 2.31 & 2.81 & 2.56   \\
            General & Webcam & 2.41 & {\bf 2.50} & 2.46    \\ 
            \hline
            \end{tabular}
            }
    \end{subtable}
    \hfill
    \vspace{10pt}
    \begin{subtable}[h]{0.48\textwidth}
        \centering
        \caption{RR}
        \scalebox{0.9}{
        \begin{tabular}{cc|ccc}  \hline
        Model & Test camera & Session 1 & Session 2 & Mean \\ \hline
        Smartphone & Smartphone & {\bf 3.62} & {\bf 5.83} & {\bf 4.73}  \\
        Webcam & Smartphone & 3.75 & 5.96 & 4.86    \\
        General & Smartphone & 3.99 & 6.04 & 5.02   \\ 
        \hline
        Webcam & Webcam & 4.14 & {\bf 3.97} & 4.06    \\
        Smartphone & Webcam & {\bf 4.05} & 3.98 & {\bf 4.02}   \\
        General & Webcam &  4.32 &  4.42 & 4.37\\ 
        \hline
        \end{tabular}
        }
     \end{subtable}
     \label{table:cross-camera}
\end{table}

Although the experimental results are promising, CalibrationPhys still has some limitations.
First, our chest motion-based RR measurement method is sensitive to noise due to camera shake on a smartphone.
In future work, we need to consider approaches to reduce camera shake or to take RR measurements based on sources other than chest motion.
Smartphones equipped with motion compensation software may help reduce the performance degradation due to camera shake.
Second, this study examined self-collected datasets, which include mainly Japanese subjects, and the UBFC-rPPG, which includes mainly Caucasian subjects.
Since it has been reported that HR estimation performance in subjects with darker skin is lower than in subjects with lighter skin~\cite{liu2021metaphys}, we should examine different racial groups in the future.
Third, the subjects were seated in this experiment, but standing, walking, or running could be considered.
Long-term monitoring is also important for the diagnosis of disease and should be explored in the future.
Furthermore, we should consider not only adults but also children and elderly patients as subjects in our experiments.

\section{Conclusion}
\label{sec:conclusions}
In this paper, we presented CalibrationPhys, a self-supervised video-based HR and RR measurement method that calibrates between multiple cameras.
CalibrationPhys does not require supervised labels for model training and enables highly accurate HR and RR measurements.
In addition, it expands the limited training data via temporal augmentation and successfully leverages a pre-trained model for a particular camera.
We investigated the differences in estimation performance between the webcam and the smartphone and confirmed that CalibrationPhys performs better than state-of-the-art methods.
Our approach requires only videos from multiple cameras for model training, thus facilitating the use of arbitrary cameras.

Traditional contact-based sensors for measuring HR and RR, such as PPG and respiratory belts, are costly to measure and require specialized medical equipment.
Our video-based measurement method can measure HR and RR on the subject's smartphone or tablet, and even remotely.
Furthermore, our method constructs camera-specific models without supervised labels, which are difficult to collect when building DNNs for medical applications~\cite{zhang2018learning}.
Our self-supervised framework may be applied not only to video-based HR and RR measurements but also to a wide range of healthcare and medical domains where collecting supervised labels is challenging.

\section*{Appendix A}
Table~\ref{table:num_frame} shows the mean absolute error (MAE) of the HR and RR measurements corresponding to several numbers of frames $T$ used for model training.
The MAE for all tasks in each session and their means
are shown.
In our experiments, $T$ was set to 150 and 300 for HR and RR measurements, respectively.
When $T$ is set too small, it becomes difficult to extract HR and RR from the waves and estimation performance degrades rapidly.
On the other hand, when $T$ is set to large values, the changes in HR and RR within $T$ frames increase, and the PSDs cannot be estimated well, resulting in gradually degrading estimation performance.

\section*{Appendix B}
We compared the estimation performance for HR and RR by changing the backbone in CalibrationPhys from our model (PhysNet-based 2DCNN) to several CNN networks.
As common CNN networks, we used ResNet18~\cite{he2016deep}, Inception-ResNet-v2~\cite{szegedy2017inception}, and Inception-v4~\cite{szegedy2017inception}, which were initialized with ImageNet~\cite{deng2009imagenet} pre-training.
Table~\ref{table:cnn} shows the MAE of the HR and RR measurements.
The MAEs of ResNet18, Inception-ResNet-v2, and Inception-v4 are much worse compared to our model.
This is because common CNN networks are not designed to output pulse or respiratory waves.
On the other hand, our PhysNet-based 2DCNN is designed to output pulse or respiratory waves from a spatio-temporal representation, and performs much better than common CNN networks.

\section*{Appendix C}
Table~\ref{table:webcam} shows the MAE for the HR and RR estimation on the webcam.
As on the smartphone, CalibrationPhys outperforms state-of-the-art methods for HR and RR measurements on the webcam as well.
Therefore, CalibrationPhys achieves the best estimation performance on both cameras used for contrastive learning.

\begin{table}[t]
     \caption{MAE for HR and RR estimation corresponding to several numbers of frames $T$ used for model training in the multi-camera dataset.}
    \begin{subtable}[h]{0.45\textwidth}
        \centering
        \caption{HR}
        \scalebox{0.9}{
        \begin{tabular}{l|ccc}  \hline
        $T$ (second) &  Session 1 & Session 2 & Mean \\ \hline
        100 (3.3s) & 11.10 & 10.21 & 10.66  \\
        {\bf 150 (5.0s)} & {\bf 4.00} & {\bf 4.55} & {\bf 4.28}  \\
        200 (6.7s) & 4.27 & 5.03 & 4.65  \\
        250 (8.3s) & 4.39 & 5.54 & 4.97  \\
        \hline
        \end{tabular}
        }
    \end{subtable}
    \hfill
    \vspace{10pt}
    \begin{subtable}[h]{0.45\textwidth}
        \centering
        \caption{RR}
        \scalebox{0.9}{
        \begin{tabular}{l|ccc}  \hline
        $T$ (second) &  Session 1 & Session 2 & Mean \\ \hline
        200 (6.7s) & 5.22 & 6.42 & 5.82  \\
        250 (8.3s) & 3.84 & \bf{5.82} & 4.83  \\
        \bf{300 (10.0s)} & \bf{3.62} & 5.83 & \bf{4.73}  \\
        350 (11.7s) & 4.03 & 6.43 & 5.23  \\
        \hline
        \end{tabular}
        }
     \end{subtable}
     \label{table:num_frame}
\end{table}

\begin{table}[t]
     \caption{MAE for HR and RR estimation on the smartphone in the multi-camera dataset using several backbones.}
    \begin{subtable}[h]{0.45\textwidth}
        \centering
        \caption{HR}
        \scalebox{0.9}{
        \begin{tabular}{l|ccc}  \hline
        Backbone &  Session 1 & Session 2 & Mean \\ \hline
        Resnet18~\cite{he2016deep} & 16.04 & 16.66 & 16.35  \\
        Inception-ResNet-v2~\cite{szegedy2017inception} & 13.41 & 13.99 & 13.70  \\
        Inception-v4~\cite{szegedy2017inception} & 16.57 & 16.75 & 16.66  \\
        {\bf Ours (PhysNet-based)} & {\bf 4.00} & {\bf 4.55} & {\bf 4.28}  \\
        \hline
        \end{tabular}
        }
    \end{subtable}
    \hfill
    \vspace{10pt}
    \begin{subtable}[h]{0.45\textwidth}
        \centering
        \caption{RR}
        \scalebox{0.9}{
        \begin{tabular}{l|ccc}  \hline
        Backbone &  Session 1 & Session 2 & Mean \\ \hline
        Resnet18~\cite{he2016deep} & 12.68 & 13.00 & 12.84  \\
        Inception-ResNet-v2~\cite{szegedy2017inception} & 12.63 & 13.18 & 12.91  \\
        Inception-v4~\cite{szegedy2017inception} & 18.64 & 17.35 & 18.00  \\
        {\bf Ours (PhysNet-based)} & \bf{3.62} & \bf{5.83} & \bf{4.73}  \\
        \hline
        \end{tabular}
        }
     \end{subtable}
     \label{table:cnn}
\end{table}

\begin{table}[t]
     \caption{MAE for HR and RR estimation on the webcam in the multi-camera dataset.}
    \begin{subtable}[h]{0.45\textwidth}
        \centering
        \caption{HR}
        \scalebox{0.9}{
        \begin{tabular}{l|ccc}  \hline
        Method &  Session 1 & Session 2 & Mean \\ \hline \hline
        \textit{Signal processing}: & & &   \\
        ICA~\cite{poh2010non} & 12.02 & 8.99 & 10.50  \\
        CHROM~\cite{de2013robust} & 3.57 & 3.90 & 3.74  \\
        POS~\cite{wang2016algorithmic} & 3.85 & 4.07 & 3.96  \\ \hline
        \textit{Supervised DL}: & & &   \\
        DeepPhys~\cite{chen2018deepphys} & 5.59 & 5.01 & 5.30  \\
        PhysNet~\cite{yu2019remote} & 2.55 & 2.94 & 2.75  \\
        TS-CAN~\cite{liu2020multi} & 3.14 & 3.57 & 3.36  \\
        EfficientPhys-C~\cite{liu2023efficientphys} & 2.83 & 2.89 & 2.86  \\ \hline
        \textit{Self-supervised DL}: & & & \\
        Contrast-Phys~\cite{sun2022contrast} & 3.92 & 4.98 & 4.45  \\
        \bf{CalibrationPhys} & \bf{2.28} & \bf{2.53} & \bf{2.41}  \\
        \hline
        \end{tabular}
        }
    \end{subtable}
    \hfill
    \vspace{10pt}
    \begin{subtable}[h]{0.45\textwidth}
        \centering
        \caption{RR}
        \scalebox{0.9}{
        \begin{tabular}{l|ccc}  \hline
        Method &  Session 1 & Session 2 & Mean \\ \hline \hline
        \textit{Benchmark}: & & & \\
        RGB & 5.07 & 4.96 & 5.02  \\
        Optical Flow & 4.62 & 4.50 & 4.56  \\ \hline
        \textit{Supervised DL}: & & &   \\
        DeepPhys~\cite{chen2018deepphys} & 6.02 & 6.40 & 6.21  \\
        TS-CAN~\cite{liu2020multi} & 5.63 & 6.04 & 5.84  \\ \hline
        \textit{Self-supervised DL}: & & & \\
        \bf{CalibrationPhys} & \bf{4.14} & \bf{3.97} & \bf{4.06}  \\
        \hline
        \end{tabular}
        }
     \end{subtable}
     \label{table:webcam}
\end{table}

\bibliographystyle{IEEEbib}
\bibliography{main}

\end{document}